\documentclass[11pt]{article}
\usepackage{hyperref}
\usepackage{amssymb}
\usepackage{graphicx}
\usepackage{amsmath}
\usepackage{authblk}
\usepackage[section]{placeins}
\usepackage[a4paper, total={6.8in, 10in}]{geometry}
\numberwithin{equation}{section}
\title{Maximal acceleration in non-commutative space-time and its implications}
\author {E. Harikumar \thanks{harisp@uohyd.ernet.in} and Vishnu Rajagopal \thanks{vishnurajagopal.anayath@gmail.com}}
\affil{School of Physics, University of Hyderabad, \\Central University P.O, Hyderabad-500046, Telangana, India}
\date{}
\begin{document}
\maketitle
\begin{abstract}
In this paper, we derive the non-commutative corrections to the maximal acceleration of a massive particle. Using the eight-dimensional $\kappa$-deformed phase-space metric, we obtain the $\kappa$-deformed maximal acceleration, valid up to first order in the deformation parameter. We then derive the $\kappa$-deformed geodesic equation and obtain its Newtonian limit and from this obtain a bound on the deformation parameter. After re-expressing the $\kappa$-deformed Schwarzschild metric in terms of maximal acceleration, we analyse the motion of a particle in this space-time, and also study the modifications to Hawking radiation. We also derive the $\kappa$-deformed corrections to maximal acceleration using $\kappa$-deformed generalised uncertainty principle.
\end{abstract}
\section{Introduction}
The study of gravity at microscale is of great interest and many quantum gravity models are being introduced to construct a renormalisable theory of quantum gravity. Each of these quantum gravity models, such as string 
theory \cite{sieberg}, loop quantum gravity \cite{cr}, asymptotic safe gravity, dynamical triangulation \cite{jurk} etc, predict the existence of a fundamental length scale associated with it\cite{connes}.

Two of the well studied non-commutative space-times are Moyal space-time and $\kappa$- space-time. In Moyal space-time, the commutation relation between space-time coordinates is a constant tensor \cite{sieberg}, i. e., 
\begin{equation*}
 [\hat{x}_{\mu},\hat{x}_{\nu}]=\Theta_{\mu\nu}.
\end{equation*}
In $\kappa$ space-time the space and time coordinates satisfy a Lie algebraic type commutation relation
\begin{equation*}
 [\hat{x}_0,\hat{x}_j]=ia\hat{x}_j,~[\hat{x}_i,\hat{x}_j]=0.
\end{equation*}
In order to treat the fundamental length-scale as a frame independent quantity, the Special Theory of Relativity is extended to Deformed/Doubly Special Relativity (DSR) \cite{kow}. The $\kappa$-Poincare algebra
incorporates the features of a DSR theory and hence the $\kappa$-deformed space-time is one among the many possible candidates of DSR space-time. The symmetry associated with the $\kappa$-space-time 
is defined using 
$\kappa$-Poincare algebra, which is a Hopf algebra \cite{majid,hopf,wjk}. It has been shown in \cite{kreal,mel1,mel2,mel3,mel4} that one can realise the symmetry algebra of $\kappa$-space-time whose defining 
relations are same as the usual Poincare algebra, but the explicit form of the generators are deformed. This algebra is known as the undeformed $\kappa$-Poincare algebra. The commutative derivatives do not transform as a 
4-vector under the undeformed $\kappa$-Poincare algebra, and in \cite{kreal} a Dirac derivative was introduced which transform as a $4$-vector. This Dirac derivative is used to construct the quadratic Casimir in 
$\kappa$-space-time \cite{kreal}.  

Different aspects of the $\kappa$-space-time has been studied in various contexts \cite{mel1,mel2,mel3,luk,mel5,pach,mel7}. In \cite{trg1}, using the underlying Hopf algebra structure, it was shown that the 
quantisation of $\kappa$-deformed Klein-Gordon equation results in the deformed oscillator algebra. In \cite{vishnu1} quantisation of then $\kappa$-deformed Klein-Gordon field was studied, without any 
reference to Lagrangian, leading to the deformed oscillator algebra. The $\kappa$-deformed Maxwell's equations were derived in \cite{juric,hari}. The $\kappa$-deformed geodesic equation was obtained 
in\cite{hjm}. In\cite{sivakumar}, the $\kappa$-deformed Dirac equation was derived from the generators of the undeformed $\kappa$-Poincare algebra. In \cite{kapoor}, $\kappa$-deformed Newton's law was derived 
from the $\kappa$-deformed Hamiltonian. The $\kappa$-deformed Casimir force was derived by using the relation between Energy-Momentum tensor and the Green's function corresponding to $\kappa$-deformed 
scalar field in the presence of two parallel plates \cite{suman}. In \cite{gauge} different gauge theories in $\kappa$-space-time were studied.

In certain quantum gravity models, the existence of a minimal length scale is shown to be associated with the existence of an upper limit on the proper acceleration, known as maximal acceleration \cite{dk}. The existence of maximal acceleration, which is compatible with the local Lorentz symmetry, in covariant loop quantum gravity has been shown in \cite{cr2}. In \cite{rp} it has been shown that an upper bound on the string acceleration, related to the string tension and mass exist. It has been shown that maximal acceleration arises when quantum mechanical considerations are incorporated into the eight-dimensional phase-space of a massive particle moving in $3+1$ dimensional space-time \cite{cai1}. The maximal acceleration has also been obtained as a consequence of the Heisenberg's uncertainty relation \cite{heis,heis1}. The maximal acceleration is also obtained by analysing the relation between temperature of the vacuum radiation in a uniformly accelerated frame to the absolute maximum temperature of the radiation \cite{brandt1}. The study of maximal acceleration in compact stars showed that maximal 
acceleration alters their stability conditions, \cite{papini}. In \cite{vvn} the maximal acceleration has been used efficiently to smoothen the UV divergences in local QFT.

In \cite{scarp1} the relativistic kinematics of a particle with maximal acceleration has been studied and new transformation law, leaving the $8$-dimensional line element invariant has been derived. The dynamics of a 
relativistic particle having maximal acceleration has been studied in various backgrounds such as Schwarzschild, Reissner-Nordstrom and Kerr \cite{back}.  It had been shown that the corrections involving Maximal 
acceleration lead to novel potential barrier, which lie out side the horizon and thus affect the dynamics of a particle falling radially inward. These analyses are done by first re-expressing the phase space 
metric as a function of maximal acceleration. The dynamics of model with maximal acceleration has been used to describe the Unruh effect and Hawking radiation, without using the quantised scalar fields and 
Bogoliubov coefficients \cite{feo}. The modified Unruh temperature was obtained from modified Rindler metric embedded with maximal acceleration \cite{elmo}.

The phenomenological consequences of model with maximal acceleration was studied in \cite{cain}. In \cite{kuwata,gs1} an upper limit on the mass of the Higgs boson was obtained by modifying the Higgs-fermion interaction due to the maximal acceleration. The existence of maximal acceleration modify the standard model of cosmology that avoids initial singularity and introduces an inflationary expansion \cite{gasp}. The maximal acceleration was shown to modify the Rindler metric and this induces a scalar curvature \cite{rind}. It has been shown that in the classical limit of $\kappa$-Poincare algebra, the acceleration of particle has a finite maximum \cite{kalyana}. Thus it is interesting to study how the maximal acceleration is modified in $\kappa$-deformed space-time and also analyze its implications. We address these issues here.

In this paper, we derive the $\kappa$-deformed corrections to maximal acceleration using the eight-dimensional $\kappa$-deformed phase-space metric. Using the deformed dispersion relation and the $\kappa$-deformed metric, the eight-dimensional phase-space metric is derived. In general, the quadratic Casimir for the undeformed $\kappa$-Poincare algebra contains infinitely many higher-order derivatives and these modifications find the way in to the phase-space metric. Here we consider only the first-order corrections due to the $\kappa$-deformation. We also derive the $\kappa$-deformed correction to the maximal acceleration using $\kappa$-deformed generalised uncertainty relation. We show that the $\kappa$-deformed correction term in both the cases depends linearly on the mass of the test particle and differs only by a numerical factor. Further, we show that the $\kappa$-deformed maximal acceleration leads to correction to the radial component of the Newton's force equation. We analyse the motion of a particle in the $\kappa$-deformed Schwarzschild space-time and show that the time taken by particle to reach the horizon gets affected by a factor that depends on the deformed maximal acceleration, energy scale of the deformed metric and mass of the incoming particle. We then show that the Hawking radiation also picks up a 
correction due to the $\kappa$-deformation. 
 
This paper is organised in the following way. In sec. 2, we derive the $\kappa$-deformed metric, valid up to first order in $a$. By taking the direct sum of the $\kappa$-deformed metric and the $\kappa$-deformed dispersion relation, an eight-dimensional metric in $\kappa$-space-time is constructed, which is valid up to first order in $a$. In sec. 3, we derive an expression for the $\kappa$-deformed maximal acceleration, valid up to first-order in $a$, from the time like line element defined in the $\kappa$-deformed eight-dimensional phase-space. We compare the maximal acceleration for electron and proton as a function of the deformation parameter. In subsec. 3.1, we derive the $\kappa$-deformed geodesic equation using the $\kappa$-deformed metric containing maximal acceleration term in it and then obtain its Newtonian limit, valid up to first order in $a$. Using the Pioneer anomaly, from the $\kappa$-deformed Newton's force law, a bound on the deformation parameter is obtained. In sec. 4, we construct the $\kappa$-deformed Schwarzschild metric, containing maximal acceleration term in it and calculate the time taken by a massive particle to reach the horizon. Further, we evaluate the gravitational redshift in the $\kappa$-space-time. In subsec. 4.1, $\kappa$-deformed corrections to the Hawking radiation is derived from the $\kappa$-deformed Schwarzschild metric. We also obtain the $\kappa$-deformed entropy and specific heat of the $\kappa$-deformed Schwarzschild black hole, valid up to first order in $a$. In sec. 5 we derive $\kappa$-deformed corrections to the maximal acceleration using the $\kappa$-deformed generalised uncertainty relations and compare it with former expression. In sec. 6 we summarise our results and give conclusions. 

Here we use $\eta_{\mu\nu}=diag(-1,1,1,1)$.

\section{$\kappa$-deformed metric in phase-space}

In this section we first give a brief summary of the construction of the $\kappa$-deformed metric in the phase-space, valid up to first order in $a$.

The $\kappa$-deformed space-time is an example for a non-commutative space-time where space coordinates commute among themselves while the commutators between time and space coordinate is proportional to space coordinate itself, i.e,
\begin{equation} \label{lie}
 [\hat{x}_i, \hat{x}_j]=0,~~[\hat{x}_0,\hat{x}_i]=ia\hat{x}_i,
\end{equation}
where $a$ is the deformation parameter having the dimension of length.

It is known that the $\kappa$-space-time is the space-time associated with doubly (or deformed) special relativity (DSR) \cite{kow,majid}. DSR was introduced to incorporate fundamental length scale that appears in various 
approaches to microscopic theory of gravity in such a way that the new relativity principle will accommodate two fundamental constants, namely velocity of light $c$ and a constant having dimensions of length 
$a=\frac{1}{\kappa}$.  Thus in DSR and hence in $\kappa$-space-time, velocity of light is a constant and it is the maximal velocity a particle can achieve (this will play an important role in the calculation of maximal 
acceleration as seen below). In the limit where the deformation parameter vanishes, i.e., $a\to 0$, DSR reduces to special theory of relativity.

In this paper, following \cite{kreal}, the symmetry algebra of the $\kappa$-deformed space-time is defined by the undeformed $\kappa$-Poincare algebra. Under this algebra the usual commutative derivatives do not 
transform as four-vectors, instead a Dirac derivative is defined to serve this purpose. The explicit form of the components of Dirac derivtaive are
\begin{equation}\label{dirac}
D_{i}=\partial_{i},~~
D_{0}=\partial_{0}\frac{\textnormal{sinh}A}{A}-\frac{i}{2}{a\partial_{i} ^2} e^{A}.
\end{equation}
where $A=-ia\partial_0$.  The undeformed $\kappa$-Poincare algebra is given by \cite{kreal}
\begin{equation}\label{palg1}
\begin{split}
 [\hat{M}_{\mu\nu},D_{\lambda}]&=\eta_{\nu\lambda}D_{\mu}-\eta_{\mu\lambda}D_{\nu},\\
[D_{\mu},D_{\nu}]&=0,\\
[\hat{M}_{\mu\nu},\hat{M}_{\lambda\rho}]&=\eta_{\mu\rho}\hat{M}_{\nu\lambda}+\eta_{\nu\lambda}\hat{M}_{\mu\rho}-\eta_{\nu\rho}\hat{M}_{\mu\lambda}-\eta_{\mu\lambda}\hat{M}_{\nu\rho}.
\end{split}
\end{equation}
Note that the $\hat{M}_{\mu\nu}$ in the above algebra, is defined as 
\begin{equation}\label{genr}
 \hat{M}_{\mu\nu}=\left(\hat{x}_{\mu}D_{\nu}-\hat{x}_{\nu}D_{\mu}\right)Z,
\end{equation}
where $Z^{-1}=iaD_0+\sqrt{1+a^2D_{\alpha}D^{\alpha}}$. Using the Dirac derivative, the quadratic Casimir of the undeformed $\kappa$-Poincare algebra is defined as 
\begin{equation}
D_{\mu}D^{\mu}=\partial_{i}^2-\left(\partial_0 \frac{sinhA}{A}-a\frac{i}{2}\partial_{i}^2e^A\right)^2
\end{equation}
which in the commutative limit reduces to the usual Casimir of the Poincare algebra. Thus using the quadratic Casimir, the $\kappa$-deformed dispersion relation is given as
\begin{equation}\label{dis}
-p_{i}^2+\frac{1}{a^2}sinh^2(ap_0)-p_{i}^2sinh(ap_0)e^{ap_0}+\frac{a^2}{2}p_{i}^4e^{2ap_0}-m^2=0,
\end{equation}
which is written in terms of the commutative variables and the deformation parameter. 

The $\kappa$-deformed metric is constructed by introducing the generalised commutation relations between the phase space coordinates \cite{hjm} as
\begin{equation}\label{phase}
 [\hat{x}_{\mu},\hat{P}_{\nu}]=i\hat{g}_{\mu\nu}(\hat{x}),
  \end{equation} 
where $\hat{g}_{\mu\nu}(\hat{x})$ is the $\kappa$-deformed metric. Using $\hat{x}_{\mu}$ and $\hat{P}_{\mu}$, one then constructs this metric explicitly. For calculational ease one introduces auxilary $\kappa$-deformed space-time coordinates $\hat{y}_{\mu}$ such that it commutes with $\hat{x}_{\mu}$. Now we consider a specifc realisation for the $\kappa$-deformed phase space coordinates \cite{kreal,hjm} given by
\begin{equation}\label{choice}
 \hat{x}_{\mu}=x_{\alpha}\varphi^{\alpha}_{\mu}, \,\hat{P}_{\mu}=g_{\alpha\beta}(\hat{y})k^{\beta}\varphi^{\alpha}_{\mu}. 
\end{equation}
Here, one assumes that $g_{\mu\nu}({\hat y})$ has same functional form as the commutative metric. This guarantees that in the commutative limit(i.e., $a\to 0$), ${\hat P}_\mu$ reduces to the correct 
commutative 4-momentum and so does Eq.(\ref{phase}). Substituting Eq.(\ref{choice}) in Eq.(\ref{lie}), we get restrictions on the function $\varphi_{\mu}^{\alpha}$ and thus we get
\begin{equation}\label{soln}
 \varphi _0^0=1, \, \varphi _i^0=0, \, \varphi_0^i=0, \, \varphi _j^i=\delta _j^i e^{-ak^0}. 
 \end{equation}
The coordinate $\hat{y}_{\mu}$ also satisfies the commutation relation similar to the one given in Eq.(\ref{lie}). As in Eq.(\ref{choice}), $\hat{y}_{\mu}$ is also written in terms of the commutative coordinates 
and their momenta, i.e, 
\begin{equation}\label{y}
 \hat{y}_{\mu}=x_{\alpha}\phi_{\mu}^{\alpha}.
 \end{equation}
Using Eq.(\ref{y}) and $[\hat{x}_{\mu},\hat{y}_{\nu}]=0$, we obtain
\begin{equation}\label{y1}
 \hat{y}_0=x_0-ax_jk^j,~~
\hat{y}_i=x_i.
 \end{equation}
Using these, the $\kappa$-deformed metric is derived \cite{zuh1,vishnu} from Eq.(\ref{phase}) and Eq.(\ref{choice}) as
\begin{equation}\label{ps}
 \hat{g}_{\mu\nu}=g_{\alpha\beta}(\hat{y})\Big(k^{\beta}\frac{\partial \varphi^{\alpha}_{\nu}}{\partial k^{\sigma}}\varphi_{\mu}^{\sigma}+\varphi_{\mu}^{\alpha}\varphi_{\nu}^{\beta}\Big). \end{equation}
Using Eq.(\ref{phase}) and Eq.(\ref{soln}) in Eq.(\ref{ps}), we obtain the explicit form of $\hat{g}_{\mu\nu}$ as \cite{zuh1}
\begin{align} \label{ps1} 
 \hat{g}_{00}&=g_{00}(\hat{y}), \\
 \hat{g}_{0i}&=g_{i0}(\hat{y})\big(1-2ak^0\big)-ag_{im}(\hat{y})k^m,\\
 \hat{g}_{i0}&=g_{i0}(\hat{y})(1-ak^0),\\
 \label{ps4}\hat{g}_{ij}&=g_{ij}(\hat{y})(1-2ak^0).
\end{align}
Since that $g_{\mu\nu}({\hat y})$ has the same functional form as the commutative metric and ${\hat y}_\mu$ is given in terms of commutative variables, above equations give the components of the non-commutative metric 
${\hat g}_{\mu\nu}$ in terms of commutative variables and the deformation parameter. Note that the modification of the metric depends on the energy scale $k^0$ \cite{zuh1}, which is coming through the realizations of 
${\hat x}_\mu$ and ${\hat y}_\mu$ given above.

The line element in $\kappa$-deformed space-time, defined  as \cite{zuh1}
\begin{equation}d\hat{s}^2=\hat{g}_{\mu\nu}d\hat{x}^{\mu}d\hat{x}^{\nu}, \end{equation}
is explicitly given, up to first order in $a$ by
\begin{equation}\label{dmetric}
\begin{split}
 d\hat{s}^2=&g_{00}(\hat{y})dx^0dx^0+g_{i0}(\hat{y})(1-3ak^0)dx^idx^0-ag_{im}(\hat{y})k^mdx^0dx^i+g_{i0}(\hat{y})(1-2ak^0)dx^idx^0\\+&g_{ij}(\hat{y})(1-4ak^0)dx^idx^j.
\end{split}
\end{equation}
and using
\begin{equation}
 g_{\mu\nu}(\hat{y}_0)=g_{\mu\nu}(x_0)-ax_jk^j\frac{\partial g_{\mu\nu}(x_0)}{\partial x_0},~~g_{\mu\nu}(\hat{y}_i)=g_{\mu\nu}(x_i),
\end{equation}
obtained by Taylor expansion of $g_{\mu\nu}(\hat{y})$ in Eq.(\ref{dmetric}) and setting $g_{\mu\nu}(x)=\eta_{\mu\nu}$, the line element in $\kappa$-deformed Minkowski space-time, valid up to first order in 
$a$ becomes ( we set $c=1$)
\begin{equation}\label{min}
 d\hat{s}^2=-dt^2+(1-4ak^0)dx^2. 
\end{equation}
Note that the line element is written in terms of commutative coordinates( i.e., coordinates of Minkowski space-time). This is due to the use of realizations of ${\hat x}_\mu$ and ${\hat y}_\mu$ in terms of commutative 
variables, which depend on the deformation parameter $a$ as well as on the energy scale $k^0$.

The $\kappa$-deformed dispersion relation Eq.(\ref{dis}), valid up to first order in $a$, is given by 
\begin{equation}\label{kdis}
 E^2-p^2(1+aE)-m^2=0.
\end{equation}
Using this, we construct an $8$-dimensional $\kappa$-deformed flat metric, valid up to first order in $a$. This is constructed by taking the direct sum of $\kappa$-deformed space-time metric and $\kappa$-deformed momentum-space metric (using Eq.(\ref{min}) and Eq.(\ref{kdis})) as \cite{cai1}
\begin{equation}\label{pp}
 d\hat{s}^2=-dt^2+(1-4ak^0)dx^2+\frac{1}{\mu^4}\Big(-dE^2+dp^2(1+aE)+ap_jdp^jdE\Big).
\end{equation}
Here $\mu$ is a parameter having the dimension of mass. Note that all the space-time coordinates appearing in the above equation are commutative ones, i.e., of Minkowski space-time.  Note here that the $\kappa$-deformed dispersion relation in Eq.(\ref{dis}) is valid to all orders in the deformation parameter $a$ and it is, by construction, invariant under the action of undeformed $\kappa$-Poincare algebra. Thus the 8-dimensional, $\kappa$-deformed line element defined in the phase space is invariant under the action of undeformed $\kappa$-Poincare algebra (valid up to first order in $a$). In the commutative limit, i.e., $a\to 0$, above line element reduces to the commutative line element \cite{cai1}.

\section{$\kappa$-deformed maximal acceleration}

In this section we calculate the $\kappa$-deformed correction to maximal acceleration, valid up to first order in $a$, using the $\kappa$-deformed 8-dimensional metric derived in the previous section. We further study the dynamics of a particle moving in this back ground by setting up geodesic equation. We also obtain its Newtonian limit and analyse how the force equation is modified by maximal acceleration introduced by the non-commutativity of the 
space-time.

The maximal acceleration for a particle in the commutative space-time had been derived from the causally connected $8$-dimensional metric \cite{cai1}, which is constructed by taking the direct sum of the flat space-time metric and the momentum-space metric. Now we use this approach for calculating the non-commutative corrections to the maximal acceleration in $\kappa$-deformed space-time, valid up to first order in $a$. Note that in the previous section
the generators of undeformed $\kappa$-Poincare algebra, its Casimir as well as the $\kappa$-deformed metric/line element all are written in terms of the commutative variables, energy scale $k^0$ and the deformation parameter.

For the time-like events, we have $d\hat{s}^2\leq 0$ (it is to be noted that for a metric with signature $\eta_{\mu\nu}=diag(1,-1,-1,-1)$, the time-like events have $d\hat{s}^2\geq 0$) and generalising this to $\kappa$-deformed case, we have
\begin{equation}
 -dt^2+(1-4ak^0)dx^2-\frac{1}{\mu^4}\Big(dE^2-(1+aE)dp^2-ap_jdp^jdE\Big)\leq 0.
\end{equation}
Dividing the above expression throughout by $dt^2$ and denoting $v=\frac{dx}{dt}$, which is the velocity of the particle, we have
 \begin{equation}\label{ds0}
 1-(1-4ak^0)v^2+\frac{1}{\mu^4}\bigg[\Big(\frac{dE}{dt}\Big)^2-(1+aE)\Big(\frac{dp}{dt}\Big)^2-ap_j\frac{dp^j}{dt}\frac{dE}{dt}\bigg]\geq 0
\end{equation}
From the $\kappa$-deformed energy-momentum dispersion relation given in Eq.(\ref{kdis}), we get $\frac{dE}{dt}=(1+aE+\frac{ap^2}{2E})\frac{p}{E}\frac{dp}{dt}$, which is valid up to first order in $a$, and we identify proper acceleration of the particle as $A$, where $\frac{dp}{dt}=\frac{mA}{(1-v^2)^{3/2}}$. Since the 3-momentum appearing in the above(which comes from Eq.(\ref{kdis})) is commutative momentum, we do not get any a-dependent correction to $\frac{dp}{dt}$. Using these definitions in the above equation, we get
 \begin{equation}\label{ds1}
 (1-v^2)\Bigg[1+\frac{4ak^0v^2}{1-v^2}+\frac{1}{\mu^4}\frac{m^2A^2}{(1-v^2)^4}\bigg(\frac{p^2}{E^2}\Big(1+2aE+\frac{ap^2}{E}\Big)-(1+aE)-\frac{ap^2}{E}\bigg)\Bigg]\geq 0.
\end{equation}

Note that in the $\kappa$-space-time, as in the Minkowski space-time, velocity satisfies $v\le c$ and the acceleration will thus be maximum when $v<<c$. Thus to find the maximal acceleration, we use an instantaneous $\kappa$-deformed rest frame of the particle where the particle velocity vanishes, i.e., $v=0$ and hence its momentum ${\vec p}$ too.
Thus the above inequality Eq.(\ref{ds1}) becomes  
\begin{equation}
 1-\frac{m^2A_{max}^2}{\mu^4}(1+am)\geq 0
\end{equation}
We obtain the expression for the maximal acceleration in the $\kappa$-deformed space-time, up to first order in $a$, by simplifying the above inequality, as
\begin{equation}\label{A}
 \hat{A}_{max}\leq m\Big(1-\frac{am}{2}\Big).
\end{equation}
Here we have set the dimension full parameter $\mu$ as $m$, which is the rest mass of the particle. In SI units Eq.(\ref{A}) becomes
\begin{equation}\label{A1}
 \hat{A}_{max}\leq \frac{mc^3}{\hslash}\Big(1-\frac{amc}{2\hslash}\Big)
\end{equation}
We observe that the $\kappa$-deformed correction to the maximal acceleration is independent of the deformation energy ($k^0$ in Eq.(\ref{pp} and Eq.(\ref{ds1}) ). The correction term dependents only on the rest 
mass of the particle\footnote{ Note that $A_{max}$ depends on $\hbar$, even in the commutative limit where $a=0$. This dependence comes through the assumption that the lengths can be measured with arbitrary 
accuracy, see\cite{cai1}.}. The deformed maximal acceleration is positive when $\frac{amc}{2\hslash}<1$ and it is negative when $\frac{amc}{2\hslash}>1$. In the commutative limit, i.e $a\to 0$, the maximal acceleration becomes $A_{\max}=\frac{mc^3}{\hslash}$ as in \cite{cai1,heis} (note that in the natural units, $\hat{A}_{max}=m$ in the commutative limit). When we go to classical limit, i.e $\hslash \to 0$, the deformed maximal acceleration blows up to infinity, as seen in \cite{cai1,heis}. We see that for a massless particle, the deformed maximal acceleration reduces to zero. 

Using the reduced Compton wavelength ($\lambda=\frac{\hslash}{mc}$) for a particle of rest mass $m$, we re-express the deformed maximal acceleration as
\begin{equation}
 \hat{A}_{max}\leq \frac{c^2}{\lambda}\Big(1-\frac{a}{2\lambda}\Big).
\end{equation}  
Using Eq.(\ref{A1}) we observe that the deformed maximal acceleration for both electron and proton is constant for the deformation parameter ranging from $10^{-35}$m to $10^{-14}$m, which is due to the fact that $\frac{am^2c^4}{2\hslash ^2}<<\frac{mc^3}{\hslash}$ over this range. Similarly we find that the ratio of the deformed maximal acceleration of proton to electron is around $10^{5}$ (in this range), as seen in the commutative case. 
\subsection{$\kappa$-deformed geodesic equation and its Newtonian limit}
In this subsection we derive the $\kappa$-deformed geodesic equation, in terms of $\hat{A}_{max}$ and obtain the corresponding Newtonian limit.
 
We express $\kappa$-deformed metric given in Eq.(\ref{min}), in terms of the deformed maximal acceleration by using Eq.(\ref{A}) and thus find
\begin{equation}\label{ge}
d\hat{s}^2=g_{00}(x)dx^0dx^0+g_{ij}(x)\Big[1-\frac{8k^0}{m}\Big(1-\frac{\hat{A}_{max}}{m}\Big)\Big]dx^idx^j.
 \end{equation}
Now we construct the $\kappa$-deformed geodesic equation using this deformed metric. This should be contrast with the result of \cite{hjm}, where it is derived using the Feynman's approach. Here, using the $\kappa$-deformed metric given in Eq.(\ref{ge}), we first evaluate the deformed Christoffel symbol, $\hat{\Gamma}^{\mu}_{\nu\lambda}$ and the non-vanishing ones are,
\begin{equation}\label{cs}
\begin{split}
 \hat{\Gamma}^{0}_{00}=\frac{1}{2}g^{00}\partial_0g_{00}, ~~\hat{\Gamma}^{0}_{ij}=\hat{\Gamma}^{0}_{ji}=-\frac{1}{2}\bigg[1-\frac{4k^0}{m}\Big(1-\frac{\hat{A}_{max}}{m}\Big)\bigg]g^{00}\partial_0g_{ij},~~\hat{\Gamma}^0_{i0}=\hat{\Gamma}^0_{0i}=\frac{1}{2}g^{00}\partial_ig_{00}\\
\hat{\Gamma}^{i}_{00}=\frac{1}{2}\bigg[1-\frac{4k^0}{m}\Big(1-\frac{\hat{A}_{max}}{m}\Big)\bigg]g^{ik}\partial_kg_{00},~~\hat{\Gamma}^{i}_{0j}=\hat{\Gamma}^{i}_{j0}=\frac{1}{2}\bigg[1-\frac{8k^0}{m}\Big(1-\frac{\hat{A}_{max}}{m}\Big)\bigg]g^{ik}\partial_0g_{kj}\\
\hat{\Gamma}^{i}_{jk}=\hat{\Gamma}^{i}_{kj}=\frac{1}{2}\bigg[1-\frac{8k^0}{m}\Big(1-\frac{\hat{A}_{max}}{m}\Big)\bigg]g^{il}\Big(\partial_jg_{lk}+\partial_kg_{jl}-\partial_lg_{jk}\Big).
\end{split}
\end{equation}
Using this in the geodesic equation 
\begin{equation}\label{geo}
 \frac{d^2\hat{x}^{\mu}}{d\tau ^2}+\hat{\Gamma}^{\mu}_{\nu\lambda}\frac{d\hat{x}^{\nu}}{d\tau}\frac{d\hat{x}^{\lambda}}{d\tau}=0,
\end{equation}
we derive the modification due to non-commutativity of the space-time.  From Eq.(\ref{choice}) we have $\hat{x}_0=x_0,~\hat{x}_i=x_i(1-ak^0)$ which is re-expressed in terms of $\hat{A}_{max}$ as $\hat{x}_i=x_i\Big[1-\frac{2k^0}{m}\big(1-\frac{\hat{A}_{max}}{m}\big)\Big]$. Using this and Eq.(\ref{cs}) in Eq.(\ref{geo}), we get
\begin{equation}\label{geo1}
\frac{d^2x^0}{d\tau ^2}+\frac{1}{2}g^{00}\partial_0g_{00}\frac{dx^0}{d\tau}\frac{dx^0}{d\tau}-\bigg[1-\frac{6k^0}{m}\Big(1-\frac{\hat{A}_{max}}{m}\Big)\bigg]g^{00}\partial_0g_{ij}\frac{dx^i}{d\tau}\frac{dx^j}{d\tau}+\bigg[1-\frac{2k^0}{m}\Big(1-\frac{\hat{A}_{max}}{m}\Big)\bigg]g^{00}\partial_ig_{00}\frac{dx^i}{d\tau}\frac{dx^0}{d\tau}=0
\end{equation}
and
\begin{equation}\label{geo2}
\begin{split}
&\bigg[1-\frac{2k^0}{m}\Big(1-\frac{\hat{A}_{max}}{m}\Big)\bigg]\frac{d^2x^i}{d\tau ^2}+\frac{1}{2}\bigg[1-\frac{4k^0}{m}\Big(1-\frac{\hat{A}_{max}}{m}\Big)\bigg]g^{ik}\partial_kg_{00}\frac{dx^0}{d\tau}\frac{dx^0}{d\tau}+\\&\bigg[1-\frac{10k^0}{m}\Big(1-\frac{\hat{A}_{max}}{m}\Big)\bigg]g^{ik}\partial_0g_{kj}\frac{dx^0}{d\tau}\frac{dx^j}{d\tau}+\bigg[1-\frac{12k^0}{m}\Big(1-\frac{\hat{A}_{max}}{m}\Big)\bigg]g^{il}\Big(\partial_jg_{lk}+\partial_kg_{jl}-\partial_lg_{jk}\Big)\frac{dx^k}{d\tau}\frac{dx^j}{d\tau}=0.
\end{split}
\end{equation}
As in the commutative space-time, the Newtonian limit of this equation is derived using three conditions, viz;
\begin{itemize}
\item particles are moving slowly, i.e, 
\begin{equation}
 \frac{d{x}_i}{d\tau}<<\frac{d{x}_0}{d\tau},
\end{equation}
\item gravatitational field is static, i.e,
\begin{equation}
 \frac{\partial {g}_{\mu\nu}}{\partial t}=0,
\end{equation}
\item gravitational field is weak and we use linearised deformed metric as
\begin{equation}\label{wfa}
 {g}_{\mu\nu}={\eta}_{\mu\nu}+{h}_{\mu\nu},~~~|{h}_{\mu\nu}|<<1.
\end{equation}
\end{itemize}
Using these three conditions in Eq.(\ref{geo1}) and Eq.(\ref{geo2}) and $\dot{x}_0=\frac{dt}{d\tau}$, we get 
\begin{equation}
 {\ddot{x}}_0=0\implies \dot{x}=\textnormal{constant}
\end{equation}
\begin{equation}
 \bigg[1-\frac{2k^0}{m}\Big(1-\frac{\hat{A}_{max}}{m}\Big)\bigg]\frac{d^2x_i}{d\tau ^2}+\frac{1}{2}\bigg[1-\frac{4k^0}{m}\Big(1-\frac{\hat{A}_{max}}{m}\Big)\bigg]\nabla _i{h}_{00}\dot{x}_0^2=0
\end{equation}
so we write this as,
\begin{equation}
 \bigg[1-\frac{2k^0}{m}\Big(1-\frac{\hat{A}_{max}}{m}\Big)\bigg]\frac{d^2x_i}{dt ^2}+\frac{1}{2}\bigg[1-\frac{4k^0}{m}\Big(1-\frac{\hat{A}_{max}}{m}\Big)\bigg]\nabla _i{h}_{00}=0
\end{equation}
Now we multiply the above equation by $\big[1+\frac{2k^0}{m}(1-\frac{\hat{A}_{max}}{m})\big]$ and keep up to first order term in $a$, so we get
\begin{equation}\label{h1}
 \frac{d^2{x}_i}{dt^2}+\frac{1}{2}\bigg[1-\frac{2k^0}{m}\Big(1-\frac{\hat{A}_{max}}{m}\Big)\bigg]\nabla _i{h}_{00}=0
\end{equation}
By inspecting Eq.(\ref{h1}) we find that $\hat{h}_{00}=\big[1-\frac{2k^0}{m}(1-\frac{\hat{A}_{max}}{m})\big]h_{00}$, where $h_{00}$ is given as $h_{00}=-\frac{2M}{r}$. In the commutative limit $\hat{A}_{max}\to m$ and then $\hat{h}_{00}=h_{00}$. Using this, from $\kappa$-deformed geodesic equation, we have $\frac{d^2\hat{x}_i}{dt^2}=-\frac{1}{2}\nabla_i\hat{h}_{00}$, and comparing this with the $\kappa$-deformed Newton's equation, $\hat{F}_i=m\frac{d^2\hat{x}_i}{dt^2}$, we find
\begin{equation}
 \hat{F}_i=F_i\Big[1-\frac{2k^0}{m}\big(1-\frac{\hat{A}_{max}}{m}\big)\Big].
\end{equation}  
Here $F_i=-\frac{mM}{r^2}$ is the commutative Newton's force equation. Here we observe that the non-commutative contribution to the Newtonian force has only radial component as in \cite{hjm}. We obtain a bound on the deformation parameter, by equating the additional constant acceleration, $8.5\times 10^{-10}ms^{-2}$ (that cannot be explained by Newton's law) due to Pioneer anomaly \cite{pioneer} with the correction term in the $\kappa$-deformed Newton's force equation, as $a\leq 10^{-39}m$. It is important to note that the modification to Newtonian force in the above equation depends on $a$ as well as on the energy scale $k^0$. In deriving this bound on $a$, we have taken $k^0$ to be the Planck energy. Since $k^0$ is also a unknown, it is interesting to use the bound obtained for $a$, by other methods, in the above and find the corresponding values of $k^0$.  Using 
$a\approx 10^{-34}m$ (see\cite{mel6}), we find $k^0\approx 2.35\times 10^{-7}J$, using $a\approx 10^{-22}m$(see\cite{sivakumar}), we get $k^0\approx 10^{-22}J$ and using $a\approx 10^{-49}m$(see\cite{kapoor}), we get 
$k^0\approx 10^{8}J$.

\section{$\kappa$-deformed Schwarzschild metric}

In this section, we analyse the effect of maximal acceleration induced by non-commutativity of the space-time on the  Schwarzschild metric and its implications. For this, we first re-express the Schwarzschild metric in the $\kappa$-deformed space-time, valid up to first order in the deformation parameter, as a function of maximal acceleration. To construct the $\kappa$-deformed Schwarzschild metric, we start from Eq.(\ref{ps1})-Eq.(\ref{ps4}). Note that 
$g_{\mu\nu}({\hat y})$ appearing on the RHS of these equations have the same functional form as the commutative metric and this allows us to readily write down $g_{\mu\nu}({\hat y})$. Further, from Eq.(\ref{y1}), we have ${\hat y}_i=x_i$ and recalling that the components of the Schwarzschild metric depends only on spatial co-ordiantes, we immediately find $g_{\mu\nu}({\hat y})=g_{\mu\nu}({\hat y}_i)=g_{\mu\nu}(x_i)$. We thus obtain the $\kappa$-deformed Schwarzschild metric when the deformation parameter, $a$ is re-expressed in terms of the deformed maximal acceleration, $\hat{A}_{max}$ using Eq.(\ref{A}) in Eq.(\ref{dmetric}) as,
\begin{equation}\label{schw}
 d\hat{s}^2=\bigg(1-\frac{2M}{r}\bigg)dt^2-\bigg[1-\frac{8k^0}{m}\Big(1-\frac{\hat{A}_{max}}{m}\Big)\bigg]\bigg(1-\frac{2M}{r}\bigg)^{-1}dr^2-\bigg[1-\frac{8k^0}{m}\Big(1-\frac{\hat{A}_{max}}{m}\Big)\bigg]r^2d\Omega^2
\end{equation}
Here the non-commutativity is present in $\hat{A}_{max}$ term. In the commutative limit, $\hat{A}_{max}$ becomes $m$ and the terms in the square bracket become $1$, hence we recover the commutative Schwarzschild metric. Note that the non-commutative Schwarzschild metric posses spherical symmetry as in the commutative case. We also notice that the non-commutativity does not induce any new horizons other than the $r=2M$, which is present in the commutative case. Using the above metric, we investigate the dynamics of a massive particle moving radially inward. We also evaluate the correction to Hawking temperature due to non-commutativity induced maximal acceleration.

Now we analyse the motion of a massive particle falling into the Schwarzschild black hole. Let its velocity be $\hat{v}^{\mu}=\frac{d\hat{x}^{\mu}}{ds}$ and assume that it is falling radially inward, so that $\hat{v}^2=\hat{v}^3=0$. Using Eq.(\ref{schw}), the corresponding deformed geodesic equation
\begin{equation}
 \frac{d\hat{v}^0}{ds}+\hat{g}^{00}\hat{\Gamma}_{0\nu\lambda}\hat{v}^{\nu}\hat{v}^{\lambda}=0,
\end{equation}
after simplification becomes 
\begin{equation}
 \hat{g}_{00}\frac{d\hat{v}^0}{ds}+\frac{d\hat{g}_{00}}{ds}\hat{v}^0\bigg[1-\frac{2k^0}{m}\Big(1-\frac{\hat{A}_{max}}{m}\Big)\bigg]=0.
\end{equation}
Solving this equation, we get 
\begin{equation}\label{v0}
 \hat{g}_{00}\hat{v}^0=\lambda_1\bigg(1-\frac{4Mk^0}{rm}\Big(1-\frac{\hat{A}_{max}}{m}\Big)\bigg),
\end{equation}
here $\lambda_1$ is a constant of the integration, and in the commutative limit it becomes $g_{00}v^0$. Now we use the relation $\hat{g}_{\mu\nu}\hat{v}^{\mu}\hat{v}^{\nu}=1$ and multiplying this with $\hat{g}_{00}$, we get
\begin{equation}\label{g0}
 \hat{g}_{00}\hat{g}_{00}\hat{v}^0\hat{v}^0+\hat{g}_{00}\hat{g}_{11}\hat{v}^1\hat{v}^1=\hat{g}_{00}.
\end{equation}  
From Eq.(\ref{schw}) we get
\begin{equation}\label{g1}
 \hat{g}_{00}\hat{g}_{11}=-\bigg[1-\frac{4k^0}{m}\Big(1-\frac{\hat{A}_{max}}{m}\Big)\bigg].
\end{equation}
Substituting Eq.(\ref{v0}) and Eq.(\ref{g1}) in Eq.(\ref{g0}), we get 
\begin{equation}\label{v1}
\hat{v}^1=-\Bigg(\lambda_1^2\bigg(1-\frac{8Mk^0}{rm}\Big(1-\frac{\hat{A}_{max}}{m}\Big)+\frac{4k^0}{m}\Big(1-\frac{\hat{A}_{max}}{m}\Big)-\Big(1-\frac{2M}{r}\Big)\bigg[1+\frac{2k^0}{m}\Big(1-\frac{\hat{A}_{max}}{m}\Big)\bigg]\Bigg)^{\frac{1}{2}}
\end{equation}
Note that $\hat{v}^1<0$ because the particle is falling radially inward. Using Eq.(\ref{v0}) and Eq.(\ref{v1}) we get
\begin{equation}\label{v2}
 \frac{d\hat{t}}{d\hat{r}}=\frac{\hat{v}^0}{\hat{v}^1}=-\frac{\lambda_1\bigg(1-\frac{4Mk^0}{rm}\Big(1-\frac{\hat{A}_{max}}{m}\Big)\bigg)\Big(1-\frac{2M}{r}\Big)^{-1}}{\Bigg(\lambda_1^2\bigg(1-\frac{8Mk^0}{rm}\Big(1-\frac{\hat{A}_{max}}{m}\Big)+\frac{4Mk^0}{rm}\Big(1-\frac{\hat{A}_{max}}{m}\Big)\bigg)-\Big(1-\frac{2M}{r}\Big)\bigg(1+\frac{4Mk^0}{rm}\Big(1-\frac{\hat{A}_{max}}{m}\Big)\bigg)\Bigg)^{\frac{1}{2}}}.
\end{equation}
Since the metric is singular at $r=2M$, we seek a solution near the horizon, i.e, at $r=2M+\epsilon$, such that $\epsilon<1$. Thus Eq.(\ref{v2}) becomes
\begin{equation}\label{v3}
\begin{split}
 \frac{d\hat{t}}{d\hat{r}}=\frac{\hat{v}^0}{\hat{v}^1}=&-\frac{\lambda_1\bigg(1-\frac{4Mk^0}{m(2M+\epsilon)}\Big(1-\frac{\hat{A}_{max}}{m}\Big)\bigg)\Big(1-\frac{2M}{2M+\epsilon}\Big)^{-1}}{\Bigg(\lambda_1^2\bigg(1-\frac{8Mk^0}{m(2M+\epsilon)}\Big(1-\frac{\hat{A}_{max}}{m}\Big)+\frac{4Mk^0}{m(2M+\epsilon)}\Big(1-\frac{\hat{A}_{max}}{m}\Big)\bigg)-\Big(1-\frac{2M}{2M+\epsilon}\Big)\bigg(1+\frac{4Mk^0}{m(2M+\epsilon)}\Big(1-\frac{\hat{A}_{max}}{m}\Big)\bigg)\Bigg)^{\frac{1}{2}}}\\
 =&-\bigg(1-\frac{4Mk^0}{m(2M+\epsilon)}\Big(1-\frac{\hat{A}_{max}}{m}\Big)\bigg)\bigg(1+\frac{2M}{\epsilon}\bigg)\Bigg[1-\Big(1-\frac{2M}{\epsilon}\Big)\bigg(\frac{1}{\lambda_1^2}-\frac{4k^0}{m}\Big(1-\frac{\hat{A}_{max}}{m}\Big)\Big(1+\frac{1}{\lambda_1^2}\Big)\bigg)\Bigg]^{-\frac{1}{2}}\\
=&-\bigg(1-\frac{2k^0}{m}\Big(\frac{\hat{A}_{max}}{m}-1\Big)\frac{2M}{\epsilon}\bigg)\Bigg[1+\frac{2M}{\epsilon}+\frac{1}{2\lambda_1^2}-\frac{2k^0}{m}\Big(1-\frac{\hat{A}_{max}}{m}\Big)\Big(1+\frac{1}{\lambda_1^2}\Big)\Bigg]\\
=&-\Bigg[1+\frac{2M}{\epsilon}+\frac{1}{2\lambda^2_1}-\frac{2k^0}{m}\Big(1-\frac{\hat{A}_{max}}{m}\Big)\bigg(1+\frac{2M}{\epsilon}+\frac{1}{\lambda_1^2}+\frac{1}{\lambda_1^2}\frac{M}{\epsilon}\bigg)\Bigg].
\end{split}
\end{equation}
Using $\frac{2M}{\epsilon}>>1$ and neglecting the $\frac{1}{\lambda_1^2}$ this reduces to
\begin{equation}
 \frac{d\hat{t}}{d\hat{r}}=-\frac{2M}{r-2M}\bigg(1-\frac{2k^0}{m}\Big(1-\frac{\hat{A}_{max}}{m}\Big)\bigg)
\end{equation}
We solve this differential equation and obtain the time taken by particle to reach a point near the horizon, at $r=2M+\epsilon$ as
\begin{equation}
 \hat{t}=-2M\bigg(1-\frac{2k^0}{m}\Big(1-\frac{\hat{A}_{max}}{m}\Big)\bigg)\textnormal{ln}\big(r-2M\big)
\end{equation}
As the particle approach $r=2M$, the time shoots up to infinity. Thus even in the non-commutative framework we see 
that the observer takes an infinite time to reach the event horizon. We find that 
$\hat{t}{}<0$ if $1>\frac{2k^0}{m}(1-\frac{\hat{A}_{max}}{m})>0$ and 
$\hat{t}{}>0$ if $\frac{2k^0}{m}(1-\frac{\hat{A}_{max}}{m})>1$ and 
$\hat{t}=0$ at $k^0=\frac{m}{2}(1-\frac{\hat{A}_{max}}{m})$. At this critical value, 
$k^0=\frac{m}{2}(1-\frac{\hat{A}_{max}}{m}),~\hat{t}$ becomes infinity at $r=2M$. The time interval between two 
events becomes a negative quantity, i.e $\Delta\hat{t}<0$, if $\frac{2k^0}{m}(1-\frac{\hat{A}_{max}}{m})>1$, and 
this is not a physcially acceptable condition and thus we find $\frac{2k^0}{m}(1-\frac{\hat{A}_{max}}{m})<1$. In the non-commutative framework 
the time taken by the particle to reach near the horizon get affected by a factor that depends on the deformation 
energy $k^0$, deformed maximal acceleretion and mass of the incoming particle. 

Inside the horizon i.e, $r<2M$, we calculate the time taken by particle to reach near the horizon around $r=2M$ at $r=2M-\epsilon$,  using the above steps and we obtain it as
\begin{equation}
 \hat{t}=2M\bigg(1-\frac{2k^0}{m}\Big(1-\frac{\hat{A}_{max}}{m}\Big)\bigg)\textnormal{ln}\big(r-2M\big)
\end{equation}
Inside the horizon the time taken by particle flips the sign, which is exaclty the same behaviour shown in the commutative case also. Note that the non-commutative correction also flips the same sign inside the horizon.

The Gravitational red-shift associated with the $\kappa$-deformed Schwarzschild metric containing deformed maximal acceleration is given by
\begin{equation}\label{rs}
 z=\sqrt{\frac{\hat{g}_{00}(r_1)}{\hat{g}_{00}(r_2)}}-1=\sqrt{\frac{1-\frac{2M}{r_1}}{1-\frac{2M}{r_2}}}-1.
\end{equation}
where $r_1$ is the radial distance at which the light is emitted and $r_2$ is the radial distance at which it is received. As in \cite{vishnu}, here also we find that the gravitational red-shift remains unaffected in the presence of $\kappa$-deformation, for this particular choice of realisation. This is in contrast with \cite{far}, where the gravitational red shift picks up a correction due to the presence of non-commutative term in the temporal component of the metric. 
\subsection{$\hat{A}_{max}$ corrections to Hawking radiation}
In this section we calculate the first order non-commutative corrections, induced through deformed maximal acceleration, to the Hawking radiation from the $\kappa$-deformed Schwarzschild black hole.

We begin with the Schwarzschild metric and Taylor expand it about the singular point. By setting the imaginary time, the black hole metric becomes Euclidean Schwarzschild metric. Now we compare this with the spherically symmetric metric in Euclidean space. We then avoid the singularity by appropriately choosing the periodicity of time and write the partition function using the Euclidean path integral. We obtain the Hawking temperature by taking the inverse of the periodicity of imaginary time. We repeat this procedure for a $\kappa$-deformed Schwarzschild metric and obtain the $\kappa$-deformed corrections to the Hawking temperature, valid up to first order in $a$. 

The $1-\frac{2M}{r}$ term present in the deformed Schwarzschild metric is re-defined as $f(r)=1-\frac{2M}{r}$. Using this, Eq.(\ref{schw}) becomes 
\begin{equation}
 d\hat{s}^2=f(r)dt^2-\frac{1}{f(r)}\bigg(1-\frac{8k^0}{m}\Big(1-\frac{\hat{A}_{max}}{m}\Big)\bigg)dr^2-r^2\bigg(1-\frac{8k^0}{m}\Big(1-\frac{\hat{A}_{max}}{m}\Big)\bigg)d\Omega^2.
\end{equation}
Let us take $r=r_s+\epsilon$ (where $r_s=2M$ is the Schwarzschild radius), so that we Taylor expand $f(r)$ up to $\epsilon$ term to have, $f(r)=f(r_s)+\epsilon f'(r_s)$, where $f'=\frac{df}{dr}$. Now we set $f(r_s)=0$ and thus we have 
\begin{equation}
 d\hat{s}^2=\epsilon f'(r_s)dt^2-\frac{1}{\epsilon f'(r_s)}\bigg(1-\frac{8k^0}{m}\Big(1-\frac{\hat{A}_{max}}{m}\Big)\bigg)d\epsilon ^2-(r_s+\epsilon)^2\bigg(1-\frac{8k^0}{m}\Big(1-\frac{\hat{A}_{max}}{m}\Big)\bigg)d\Omega^2.
\end{equation}
Let us introduce a new variable $\rho=\sqrt{\frac{\epsilon}{f'(r_s)}}$ and take $\tau=it$. Thus the deformed line element becomes
\begin{equation}\label{im}
 d\hat{s}^2=-\rho^2\Big(f'(r_s)\Big)^2d\tau ^2-4\bigg(1-\frac{8k^0}{m}\Big(1-\frac{\hat{A}_{max}}{m}\Big)\bigg)d\rho ^2-\Big(r_s+\rho ^2f'(r_s)\Big)^2\bigg(1-\frac{8k^0}{m}\Big(1-\frac{\hat{A}_{max}}{m}\Big)\bigg)d\Omega ^2.
\end{equation}
Now we consider the first two terms of the metric in Eq.(\ref{im}) i.e, terms corresponding to $\rho$ and $\tau$ coordinates. Thus in $\tilde{\rho}-\tau$ plane alone we have
\begin{equation}\label{im1}
 d\hat{s}^2=d\tilde{\rho}^2+\bigg[\frac{\tilde{\rho}}{2}f'(r_s)\Big(1+\frac{4k^0}{m}\big(1-\frac{\hat{A}_{max}}{m}\big)\Big)\bigg]^2d\tau ^2,
\end{equation}
where we have used the definition, $\tilde{\rho}=2\Big(1-\frac{2k^0}{m}\big(1-\frac{\hat{A}_{max}}{m}\big)\Big)\rho$. We observe that $\tilde{\rho}-\tau$ plane is flat in polar coordinates only if $|f'(r_s)|\tau$ has a period of $2\pi$, otherwise the resulting space-time will have a conical singularity at $\tilde{\rho}=0$, as result the curvature becomes infinite and we cannot extremise the action at that point. Let $\beta$ be the periodicity of $\tau$ and the area swept by $\tilde{\rho}$ in $\tilde{\rho}-\tau$ plane is $\pi r_s^2$. So we have
\begin{equation}
\int_0^{\beta}\int_0^{r_s} \frac{\tilde{\rho}}{2}|f'(r_s)|\Big(1+\frac{4k^0}{m}\big(1-\frac{\hat{A}_{max}}{m}\big)\Big) d\tilde{\rho} d\tau=\pi r_s^2.
\end{equation}
This sets the periodicity to be
\begin{equation}\label{beta}
 \beta=\frac{4\pi}{|f'(r_s)|}\Big(1-\frac{4k^0}{m}\big(1-\frac{\hat{A}_{max}}{m}\big)\Big).
\end{equation}
Using the Euclidean path integral taken over a field $\phi(\tilde{\rho},\tau)$ (where $\beta$ is the periodicity of $\tau$), we have the partition function given as
\begin{equation}
 Z=\int [D\phi]e^{-S_E[\phi]}
\end{equation} 
where $S_E[\phi]$ is the Euclidean action. If the space-time where $S_E[\phi]$ is defined does not have any conical singularity, then the system is in equilibrium and one we write the partition function as
\begin{equation}
 Z=Tr\Big(e^{-\beta H}\Big),
\end{equation}
where $H$ is Hamiltonian of the system. The equilibrium temperature $T$ is then $T=\frac{1}{\beta}$, which is called the Hawking temperature of the black hole. Thus from Eq.(\ref{beta}), we find the deformed Hawking temperature containing maximal acceleration term to be
\begin{equation}
 \hat{T}_H=\frac{1}{8\pi M}\bigg(1+\frac{4k^0}{m}\Big(1-\frac{\hat{A}_{max}}{m}\Big)\bigg).
\end{equation}
The non-commutative correction term in the Hawking temperature depends on the deformation energy, deformed maximal acceleration and mass of the particle falling into black hole.  Note that the dependence of Hawking temperature on the energy $k^0$ is coming from the dependence of the deformed metric on $k^0$ (see Eq,(\ref{ps1}-Eq.(\ref{ps4})).

In that commutative limit $\hat{A}_{max}$ becomes $m$, and we get the commutative Hawking temperature devoid of any maximal acceleration dependent terms. From the black hole thermodynamics we obtain the deformed entropy as
\begin{equation}\label{ent}
 \hat{S}=\int \frac{dM}{\hat{T}_H}=\int dM ~8\pi M\Big(1-\frac{4k^0}{m}\big(1-\frac{\hat{A}_{max}}{m}\big)\Big)=4\pi M^2\Big(1-\frac{4k^0}{m}\big(1-\frac{\hat{A}_{max}}{m}\big)\Big).
\end{equation}
For the deformed Schwarzschild black hole, we calculate the deformed specfic heat capacity as
\begin{equation}
 \hat{C}=\frac{\partial \hat{T}_H}{\partial M}=-\frac{1}{8\pi M^2}\bigg(1+\frac{4k^0}{m}\Big(1-\frac{\hat{A}_{max}}{m}\Big)\bigg)
\end{equation} 
As in the commutative case here also the specific heat is a negative quantity. The deformed Schwarzschild black hole which is in equilibrium at deformed Hawking temperature is unstable, so it absorbs the radiation and the mass of black hole increases and its temperature decreases resulting in a negative specific heat in deformed space-time.

\section{Maximal acceleration from $\kappa$-deformed uncertainty principle}

In this section we derive the $\kappa$-deformed corrections to maximal acceleration, valid up to first order in $a$, using the $\kappa$-deformed uncertainty relation between position and momenta.
  
We begin with the uncertainty relation between energy and a function of time \cite{heis,heis1} defined as, 
\begin{equation}\label{lan}
 \Delta E\Delta g(t) \geq \frac{1}{2}\frac{dg}{dt}
\end{equation}
By choosing $g=v$ and $g=x$ in Eq.(\ref{lan}) we obtain the uncertainty relation between energy and velocity as well as that between energy and position as  
\begin{equation}\label{a}
  \Delta E\Delta v(t) \geq \frac{1}{2}\frac{dv}{dt} 
\end{equation}
\begin{equation}\label{v}
 \Delta E\Delta x(t) \geq \frac{1}{2}\frac{dx}{dt} 
\end{equation}
Combining Eq.(\ref{a}) and Eq.(\ref{v}), we get
\begin{equation}
 \big(\Delta E\Delta x\big)\big(\Delta E\Delta v\big) \geq \frac{1}{4}vA
\end{equation}
where we have used the identifications, $\frac{dx}{dt}=v$ and $\frac{dv}{dt}=A$
\begin{equation}
 \big(\Delta E\big)^2\Delta x\frac{\Delta v}{v}\geq \frac{1}{4}A
\end{equation}
Now we take $\Delta E=v\Delta p$ and $v$ is defined as $v=\frac{dE}{dp}$, so we get
\begin{equation}
 \big(\Delta p\big)^2\Delta x\big({\Delta v}\big)v\geq \frac{1}{4}A.
\end{equation}
We know that the uncertainty in the velocity of the particle cannot exceed its maximum attainable velocity and from special theory of relativity this maximum attainable velocity should be less than the velocity of light, i.e, $(\Delta v)^2=<v^2>-<v>^2\leq v_{max}\leq 1$, so we take $(\Delta v)v\leq 1$. Using this relation in the above equation we get 
\begin{equation}\label{un}
 \big(\Delta p\Delta x\big)^2\geq \frac{1}{4}A\Delta x
\end{equation}
In \cite{anjana} the $\kappa$-deformed uncertainty relation between $\Delta x$ and $\Delta p$ has been derived up to first order in $a$, by imposing self-similarity condition on the path of relativistic quantum particle in $\kappa$-space-time and it is given by 
\begin{equation}\label{ku}
 \Delta x\Delta p\Big(1+\frac{a}{2\Delta x}\Big)\geq \frac{1}{2}
\end{equation}
Using Eq.(\ref{ku}) in Eq.(\ref{un}) and equating $\Delta x=\lambda$, where $\lambda$ is the reduced Compton wavelength, we obtain the $\kappa$-deformed maximal acceleration, valid up to first order in $a$, as
\begin{equation}
 \hat{A}_{max}\leq m\Big(1-am\Big)
\end{equation}
In SI units, the expression for the $\kappa$-deformed maximal acceleration takes the form
\begin{equation}\label{A2}
 \hat{A}_{max}\leq \frac{mc^3}{{\hslash}}\Big(1-\frac{amc}{{\hslash}}\Big)
\end{equation}
As in Eq.(\ref{A1}) here also the $\kappa$-deformed correction term, valid up to first order in $a$, of the maximal acceleration depends on the rest mass of the particle. The correction term of the maximal acceleration differes only by a numerical factor of $1/2$ when compared with Eq.(\ref{A1}). The commutative limit of $\hat{A}_{max}$ obtained in Eq.(\ref{A2}) is same as that in Eq.(\ref{A1}). Similarly in the classical limit, i.e $\hslash \to 0$, the $\kappa$-deformed maximal acceleration blows up to infinity, as seen in Eq.(\ref{A1}). For a massless particle, the deformed maximal acceleration reduces to zero. 
\section{Conclusions}
In this paper we have calculated the non-commutative corrections to the maximal acceleration of a massive test particle in the $\kappa$-deformed space-time. We have constructed the $\kappa$-deformed metric in phase space, by taking the direct sum of the $\kappa$-deformed flat space metric and the $\kappa$-deformed dispersion relation, valid up to first order in $a$. By demanding the corresponding line element to be causally connected, we have obtained the $\kappa$-deformed maximal acceleration, valid upto first order in $a$. The $\kappa$-deformed correction to the deformed maximal acceleretaion does not depend on the deformation energy (unlike the deformed metric), but it depends on the mass of the test particle, and it reduces to zero for a massless particle. In the commutative limit the deformed maximal acceleration becomes exaclty the same as that obtained in \cite{cai1,heis}. Similarly the deformed maximal acceleration become infinity when one approaches the classical limit, i.e $\hslash\to 0$. We observe that the $\kappa$-deformed contribution does not affect the value of maximal acceleration of electron and proton for the deformation parameter ranging from $10^{-35}$m to $10^{-14}$m. 

Using the $\kappa$-deformed metric, we have evaluated the deformed Christoffel symbol valid up to first order in $a$ and derived the $\kappa$-deformed geodesic equation, where deformation depends on the deformed maximal acceleration. Using this we obtained the corresponding Newtonian limit in the $\kappa$-space-time. We find that the correction term in the deformed Newton's force equation has only the radial component. Comparing the $\kappa$-deformed Newton's force equation with the Pioneer anomaly, we obtained a bound on the deformation parameter as $a\leq 10^{-39}$m where we have taken $k^0$ to be the Planck energy. We have also used this modification of Newton's force to find possible values of $k^0$ for different bounds of $a$ obtained in\cite{mel6, sivakumar, kapoor}. We also derived the $\kappa$-deformed Schwarzschild metric, containing deformed maximal acceleration and studied the motion of a massive particle falling radially inward. We find that this deformed Schwarzschild metric does not induce any new horizon other than the horizon located at the Schwarzschild radius. 
This should be contrasted with the results obtained in the commutative space-time in\cite{back}.  But as in the commutative case, in the $\kappa$-space-time also we see that the observer takes an infinite time to reach the event horizon, but this time gets modified by a factor that depends on the deformation energy, deformed maximal acceleration and mass of the incoming particle. Inside the horizon, it flips the sign, as seen in the commutative framework. The temporal component of the metric remains unaffected under the $\kappa$-deformation, and hence the gravitational red-shift is undeformed.

We have derived $\kappa$-deformed corrections to the Hawking temperature from the deformed Schwarzschild metric using imaginary time method. The $\kappa$-deformed correction, expressed in terms of $a$, is the same as that in \cite{zuh1}, where the $\kappa$-deformed correction to the Hawking temperature is obtained using the method of Bogoliubov coefficients. Using the deformed Hawking temperature we have derived deformed entropy and heat capacity, which involve deformed maximal acceleration. As in the commutative case here also we find that the specific heat is a negative quantity.

We also derived $\kappa$-deformed corrections, valid upto first order in $a$, to the maximal acceleration using the $\kappa$-deformed uncertainty relation. Here also the correction term depends on the rest mass of the particle as in the previous case. The correction term of the deformed maximal acceleration differs only by a numerical factor of $1/2$ as compared with Eq.(\ref{A1}). In both the cases we find the $\kappa$-deformed correction, valid up to first order in $a$, is independent of the deformation energy of the background metric. 
\subsection*{\bf Acknowledgments}
We thank the anonymous referee for useful comments and suggestions. EH dedicates this paper to the memory of Prof. M. V. George, who cared enough to make a difference to many. EH thanks 
SERB, Govt. of India, for support through EMR/2015/000622. VR thanks Govt. of India, for support through DST-INSPIRE/IF170622.

\end{document}